  \documentclass[aps,prl,twocolumn,groupedaddress,floatfix]{revtex4}
\usepackage{amssymb,amsmath,graphicx,multirow}

\begin{document}

\title{A New Direction in Dark-Matter Complementarity:\\
    Dark-Matter Decay as a Complementary Probe of Multi-Component Dark Sectors} 
\author{Keith R. Dienes$^{1,2}$\footnote{E-mail address:  {\tt dienes@email.arizona.edu}},
      Jason Kumar$^{3}$\footnote{E-mail address:  {\tt jkumar@hawaii.edu}},
      Brooks Thomas$^{4}$\footnote{E-mail address:  {\tt bthomas@physics.carleton.ca}},
      David Yaylali$^{3}$\footnote{E-mail address:  {\tt yaylali@hawaii.edu}}}
\affiliation{
     $^1\,$Department of Physics, University of Arizona, Tucson, AZ  85721  USA\\
     $^2\,$Department of Physics, University of Maryland, College Park, MD  20742  USA\\
     $^3\,$Department of Physics, University of Hawaii, Honolulu, HI 96822  USA\\
     $^4\,$Department of Physics, Carleton University, Ottawa, ON K1S 5B6 Canada}
%  \date{\today}

\begin{abstract}
In single-component theories of dark matter, the $2\protect\to 2$ amplitudes for dark-matter production, annihilation, and scattering can be related to each other through various crossing symmetries.  
These crossing relations lie at the heart of the
celebrated complementarity 
which underpins different existing dark-matter search techniques and strategies.
In {\it multi-component}\/ theories of dark matter, by contrast, there can be many different dark-matter components 
with differing masses.
This then opens up a new, ``diagonal'' direction for dark-matter complementarity: 
the possibility of dark-matter {\it decay}\/ from heavier to lighter dark-matter components. 
In this work, we discuss how this new direction may be correlated with the others,
and demonstrate that the enhanced complementarity which emerges 
can be an important ingredient 
in probing and constraining the parameter spaces of such models. 
\end{abstract}
%  \pacs{95.35.+d}

\maketitle

%========================================================================
%          KEYSROKE-SAVING MACROS, nothing complicated
%========================================================================
\newcommand{\newc}{\newcommand}
\newc{\gsim}{\lower.7ex\hbox{$\;\stackrel{\textstyle>}{\sim}\;$}}
\newc{\lsim}{\lower.7ex\hbox{$\;\stackrel{\textstyle<}{\sim}\;$}}
\makeatletter
\newcommand{\biggg}{\bBigg@{3}}
\newcommand{\Biggg}{\bBigg@{4}}
\makeatother

\def\vac#1{{\bf \{{#1}\}}}

\def\beq{\begin{equation}}
\def\eeq{\end{equation}}
\def\beqn{\begin{eqnarray}}
\def\eeqn{\end{eqnarray}}
\def\calM{{\cal M}}
\def\calV{{\cal V}}
\def\calF{{\cal F}}
\def\half{{\textstyle{1\over 2}}}
\def\quarter{{\textstyle{1\over 4}}}
\def\ie{{\it i.e.}\/}
\def\eg{{\it e.g.}\/}
\def\etc{{\it etc}.\/}

%     The following macros are to create the "blackboard bold"
%     characters for "R" (set of real numbers),
%     "C" (set of complex numbers), and "Q" (set of rational numbers).

\def\inbar{\,\vrule height1.5ex width.4pt depth0pt}
\def\IR{\relax{\rm I\kern-.18em R}}
 \font\cmss=cmss10 \font\cmsss=cmss10 at 7pt
\def\IQ{\relax{\rm I\kern-.18em Q}}
\def\IZ{\relax\ifmmode\mathchoice
 {\hbox{\cmss Z\kern-.4em Z}}{\hbox{\cmss Z\kern-.4em Z}}
 {\lower.9pt\hbox{\cmsss Z\kern-.4em Z}}
 {\lower1.2pt\hbox{\cmsss Z\kern-.4em Z}}\else{\cmss Z\kern-.4em Z}\fi}
\def\thbar{\bar{\theta}}
\def\fhatPQ{\hat{f}_{\mathrm{PQ}}}
\def\fPQ{f_{\mathrm{PQ}}}
\def\mPQ{m_{\mathrm{PQ}}}
\def\wtl{\widetilde{\lambda}}
\def\ta{\widetilde{a}}
\def\TBBN{T_{\mathrm{BBN}}}
\def\OmegaCDM{\Omega_{\mathrm{CDM}}}
\def\OmegaDM{\Omega_{\mathrm{CDM}}}
\def\Omegatot{\Omega_{\mathrm{tot}}}
\def\rhocrit{\rho_{\mathrm{crit}}}
\def\alMRE{a_{\lambda,\mathrm{M}}}
\def\aldotMRE{\dot{a}_{\lambda,\mathrm{M}}}
\def\tMRE{t_{\mathrm{MRE}}}
\def\TMRE{T_{\mathrm{MRE}}}
\def\tQCD{t_{\mathrm{QCD}}}
\def\tauMRE{\tau_{\mathrm{M}}}
\def\mPQdot{\dot{m}_{\mathrm{PQ}}}
\def\mPQddot{\ddot{m}_{\mathrm{PQ}}}
\def\mPQbar{\overline{m}_{\mathrm{PQ}}}
\def\mXdot{\dot{m}_X}
\def\mXddot{\ddot{m}_X}
\def\mXbar{\overline{m}_X}
\def\TRH{T_{\mathrm{RH}}}
\def\tRH{t_{\mathrm{RH}}}
\def\LambdaQCD{\Lambda_{\mathrm{QCD}}}
\def\fhatX{\hat{f}_X}
\def\tnow{t_{\mathrm{now}}}
\def\Omvac{\Omega_{\mathrm{vac}}^{(0)}}
\def\arcsinh{\mbox{arcsinh}}
\def\zRH{z_{\mathrm{RH}}}
\def\zMRE{z_{\mathrm{MRE}}}
\def\zinit{z_{\mathrm{init}}}
\def\tinit{t_{\mathrm{init}}}
\def\sinit{s_{\mathrm{init}}}
\def\sRH{s_{\mathrm{RH}}}
\def\sMRE{s_{\mathrm{MRE}}}
\def\snow{s_{\mathrm{now}}}
\def\BRgamma{\mathrm{BR}_{\lambda}^{(2\gamma)}}
\def\te{t_{\mathrm{early}}}
\def\tl{t_{\mathrm{late}}}
\def\Ehi{E_{\mathrm{high}}}
\def\Elo{E_{\mathrm{low}}}
\def\tBBN{t_{\mathrm{BBN}}}
\def\tosc{t_{\mathrm{osc}}}
\def\Tosc{T_{\mathrm{osc}}}
\def\Tnow{T_{\mathrm{now}}}
\def\Tmax{T_{\mathrm{max}}}
\def\fhatX{\hat{f}_X}
\def\LambdaG{\Lambda_G}
\def\mX{m_X}
\def\tX{t_X}
\def\tG{t_G}
\def\OmegaDM{\Omega_{\mathrm{CDM}}}
\def\ntrans{n_{\mathrm{trans}}}
\def\nosc{n_{\mathrm{osc}}}
\def\ninf{n_{\mathrm{inf}}}
\def\nG{n_{G}}
\def\ndec{n_{\mathrm{dec}}}
\def\ncut{n_{\mathrm{cut}}}
\def\nexpl{n_{\mathrm{expl}}}
\def\tLS{t_{\mathrm{LS}}}
\def\Ech{T_{\mathrm{ch}}}
\def\Omegavac{\Omega_{\mathrm{vac}}}
\def\etanow{\eta_\ast}
\def\lambdadec{\lambda_{\mathrm{dec}}}
\def\lambdatrans{\lambda_{\mathrm{trans}}}
\def\Omegatotnow{\Omega_{\mathrm{tot}}^\ast}
\def\sigmaSI{\sigma^{(\mathrm{SI})}}
\def\weff{w_{\mathrm{eff}}}
\def\vmin{v_{\mathrm{min}}}
\def\vmax{v_{\mathrm{min}}}
\def\vesc{v_{\mathrm{esc}}}
\def\erf{\mathrm{erf}}
\def\Emin{E_R^{\mathrm{min}}}
\def\Emax{E_R^{\mathrm{max}}}
\def\keVee{\mathrm{keV}_{\mathrm{ee}}}
\def\keVnr{\mathrm{keV}_{\mathrm{nr}}}
\def\Leff{\mathcal{L}_{\mathrm{eff}}}
\def\Ivint{\mathcal{I}(m_j)}
\def\rhototloc{\rho^{\mathrm{loc}}_{\mathrm{tot}}}
\def\tls{\widetilde{s}}
\def\tlp{\widetilde{p}}
\def\tlv{\widetilde{v}}
\def\tla{\widetilde{a}}
\def\tlt{\widetilde{t}}
\def\Kbar{\overline{K}}
\def\mTr{\mathrm{Tr}}
\def\tHS{t_{\mathrm{HS}}}
\def\tev{\, {\rm TeV}}
\def\gev{\, {\rm GeV}}
\def\mev{\, {\rm MeV}}
\def\kev{\, {\rm keV}}
\def\swsq{\sin^2\theta_W}
\def\xfb{\, {\rm fb}}
\def\qbar{{ \overline{q} }}
\def\Nbar{{ \overline{N} }}
\def\chibar{{ \overline{\chi} }}
\def\psibar{{ \overline{\psi} }}

\newcommand{\fb}{\rm fb}
\newcommand{\ifb}{\rm fb^{-1}}
\newcommand{\pb}{\rm pb}
\newcommand{\ipb}{\rm pb^{-1}}
\newcommand{\ssection}[1]{{\em #1.\ }}

%========================================================================

\input epsf

%========================================================================
%========================================================================
%               MAIN TEXT BEGINS HERE
%========================================================================

%========================================================================

%\tableofcontents

%%%%%%%%%%%%%%%%%%%%%%%%%%%%%%%%%%%%%%%%%%%%%%%%%%%%%%%%%%%%%%%%%%%%%%%%%%%%%%%%%%%%%%

%\section{Introduction\label{sec:introduction}}

%%%%%%%%%%%%%%%%%%%%%%%%%%%%%%%%%%%%%%%%%%%%%%%%%%%%%%%%%%%%%%%%%%%%%%%%%%%%%%%%%%%%%%

 {\it Introduction.}---  In recent years, many search techniques have been  
exploited in the hunt for dark matter~\cite{JungmanKamionkowskiGriest}.
These include possible dark-matter production 
at colliders;  direct detection of cosmological dark matter 
through its scattering off ordinary matter at underground experiments;
and indirect detection of dark matter through observation of
the remnants of the annihilation of cosmological dark matter into ordinary matter
at terrestrial or satellite-based experiments.
At first glance, these different techniques may seem to rely on
three independent properties of dark matter, namely its amplitudes
for production, scattering, and annihilation.
However, 
these three amplitudes are 
often 
related to each other through various
crossing symmetries.
As a result, the different corresponding search techniques 
are actually correlated with each other
through their dependence
on a single underlying 
interaction which couples dark matter to ordinary matter, and 
the results achieved through any one of these search techniques will have
immediate implications for the others as well as for this underlying interaction.
This is the origin of the celebrated complementarity
which connects the different existing dark-matter search techniques (for 
a review, see Ref.~\cite{complementarity}).

%==========================================================================
\begin{figure*}[t!]
\begin{center}
  \epsfxsize 6.15 truein \epsfbox {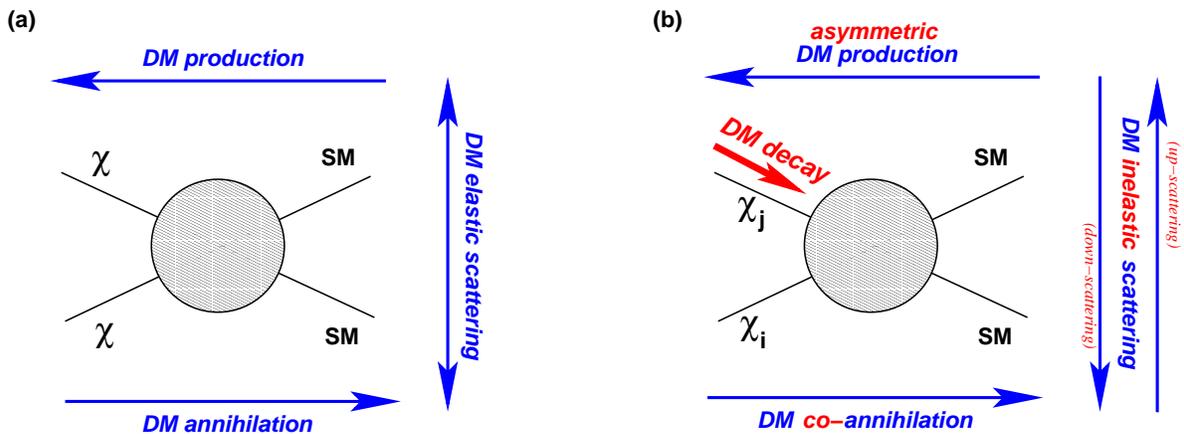}
\end{center}
\caption{(a)  In single-component theories of dark matter, the $2\protect\to 2$ amplitudes for dark-matter production, annihilation, and scattering are related to each other through various crossing symmetries.  These different
processes correspond to different directions (blue arrows) 
for the imagined flow of time through a single four-point diagram.
(b) In {\it multi-component}\/ theories of dark matter, by contrast, there can be many different dark-matter components $\chi_i$ with differing masses $m_i$.  
Taking $m_i\not = m_j$
then changes the kinematics associated with each of the previous complementary directions:
dark-matter production becomes {\it asymmetric}\/ rather than symmetric; 
dark-matter annihilation of one dark particle against itself becomes {\it co}\/-annihilation between
two different dark species;
and elastic dark-matter scattering becomes {\it inelastic}\/,
taking the form of either ``down-scattering'' or ``up-scattering''
depending on whether it is the incoming or outgoing dark-matter particle which has greater mass.
Even more importantly, however, the existence of a non-minimal dark sector
opens up the possibility for an additional process which is also related the others by crossing 
symmetries:  dark-matter {\it decay}\/ from heavier to lighter dark-matter components.
This process corresponds to a {\it diagonal}\/ direction for the imagined flow of time, as shown,
and thus represents a new direction for dark-matter complementarity.} 
\label{fig:crossings}
\end{figure*}
%==========================================================================

Most studies of this complementarity
implicitly assume that the dark sector
consists of a single particle $\chi$. 
Indeed, many studies further focus on a particular effective
$2\to 2$ interaction between dark matter and ordinary matter, 
as sketched in Fig.~\ref{fig:crossings}(a);
the three complementary search strategies then correspond to 
physical processes in which we imagine time flowing
in the directions associated with the three different blue arrows.
Indeed, because of the assumed single-particle nature of the dark sector,
we further observe that the scattering that underlies direct-detection 
experiments is necessarily elastic.
Likewise, the dark-matter-induced production of 
ordinary matter that underlies indirect-detection experiments exclusively
takes place through annihilation of the dark-matter particle
with itself or its antiparticle.

In this paper, we point out that
this situation becomes far richer  
in {\it multi-component}\/ theories of dark matter. 
In particular, if we assume that the dark sector
consists of at least two different
dark-matter components $\chi_i$ and $\chi_j$ with differing masses $m_i\not= m_j$,
then for $i\not=j$ the situation differs from
the single-particle case in two fundamental ways.
First, the kinematics associated with each of the traditional  
complementarity directions is altered:
dark-matter production becomes {\it asymmetric}\/ rather than symmetric;
dark-matter annihilation of one dark particle against itself or its antiparticle
becomes {\it co}\/-annihilation between two different dark species;
and dark-matter scattering --- previously exclusively elastic ---  now becomes {\it inelastic}\/,
taking the form of either ``up-scattering''~\cite{upscattering} or ``down-scattering''~\cite{downscattering} 
depending on whether it is the incoming or outgoing dark-matter particle which has greater mass.
These kinematic changes are illustrated in Fig.~\ref{fig:crossings}(b),
and can significantly affect the phenomenology of the corresponding processes.
But perhaps even more importantly,
an entirely new direction for dark-matter complementarity
also opens up:  this is the possibility of 
dark-matter {\it decay}\/ from heavier to lighter dark-matter components.
Indeed, this process corresponds to the {\it diagonal}\/ direction for the 
imagined flow of time, as shown in Fig.~\ref{fig:crossings}(b),
and thus represents an entirely new direction for dark-matter complementarity.
Such a direction
was not available in single-component
theories of dark matter due to phase-space constraints, and is ultimately
driven by the non-zero mass difference between the associated parent and daughter
dark-matter particles.
Of course, we are not the first to discuss decaying dark 
matter of this form (see, {\it e.g.}\/,
Refs.~\cite{decayingDM,downscattering,SMdecays} for prior work).
However our main point is that these decays are
actually part of a larger complementarity, and that this
enhanced complementarity
can be an important ingredient
in probing and constraining the
parameter spaces of theories
with non-trivial dark sectors.

Thus, 
if the dark sector consists of multiple components,
a generic four-point interaction
of the sort shown in Fig.~\ref{fig:crossings}
will lead to an enhanced set of complementarities
across different classes of dark-matter experiments.
Direct-detection experiments
will potentially be sensitive to nuclear recoils from
elastic scattering, up-scattering, and down-scattering,
while
indirect-detection experiments
will potentially be able to measure fluxes from
dark-matter self-annihilation, dark-matter co-annihilation, 
and dark-matter decay.
Likewise, collider experiments will potentially
involve dark-matter production which is both symmetric and asymmetric.
Of course, neither inelastic down-scattering nor dark-matter decay
will be relevant for present-day experiments 
unless the more massive dark-matter components
have significant cosmological abundances at the present time;
indeed, this already places one set of constraints on their decay widths.  
In this connection, we also note
that this same four-point interaction 
can in principle also give rise to an additional process
in which a heavy ordinary-matter particle 
decays to a lighter ordinary-matter particle along with two dark-sector
particles.  However, for all cases involving two quarks (which will be our 
main interest in this paper),
such processes cannot occur in a flavor-conserving theory.
Processes of this type will therefore not be considered further in this paper.  
We nevertheless note that 
in general,
exotic decays of the Higgs or of Standard-Model electroweak gauge bosons 
could potentially provide yet another complementary probe into the nature of the dark sector~\cite{SMdecays}.

%=====================================================
 {\it Two Examples.}\/--- In order to illustrate these extra complementarities and 
the power they provide in surveying the parameter space of theories with
multi-component dark sectors, let us imagine that 
the dark sector consists of two Dirac fermions $\chi_1$ and $\chi_2$ which are neutral under
all Standard-Model gauge symmetries and have corresponding masses $m_1$ and $m_2$ respectively,
with $m_2>m_1$.
We shall also assume that our fundamental four-point interaction
in Fig.~\ref{fig:crossings}(b) is described by an effective dimension-six 
four-fermi contact Lagrangian operator that couples these two dark-matter particles 
to two Standard-Model quarks $q$.
For concreteness, in this paper we shall consider two distinct examples of such operators,
with the first taking the form of a flavor-conserving scalar (S) interaction
\beq
     {\cal L}^{\rm (S)}_{\rm int} ~=~ 
            \sum_{q=u,d,s,...}
         { c_{q}^{\rm (S)}\over \Lambda^2}
               (\chibar_2 \chi_1) (\qbar q)~+~{\rm h.c.}~,
\label{scalarintpre}
\eeq
and the second taking the form of a flavor-conserving axial-vector (A) interaction
\beq
     {\cal L}^{\rm (A)}_{\rm int} ~=~ 
            \sum_{q=u,d,s,...}
         { c_{q}^{\rm (A)}\over \Lambda^2}
               (\chibar_2 \gamma_\mu \gamma^5 \chi_1) (\qbar \gamma^\mu \gamma^5 q)~+~{\rm h.c.}~~
\label{axialintpre}
\eeq
We have chosen these two forms of interactions as canonical examples
of operators giving rise to spin-independent (SI) and spin-dependent (SD) 
interactions, respectively.
In these operators, $q=u,d,s,...$ specifies a particular quark flavor
while $c_q$ is the corresponding dark-matter/quark coupling
and $\Lambda$ denotes the mass scale of the new (presumably flavor-diagonal)
physics which might generate such effective interactions.
Note that these operators are separately invariant under both charge conjugation (C) and parity (P).
Of course, the scalar operator structure explicitly violates the
chiral symmetries of the Standard Model. The coefficient of this operator thus
implicitly includes a vev of the Higgs field, so that we would in this case
more precisely identify $\Lambda' \equiv (\Lambda^2 v)^{1/3}$
as the scale of new physics.    

Within the operators in Eqs.~(\ref{scalarintpre})
and (\ref{axialintpre}),
we shall make two further assumptions.
First, in each case, we shall assume a flavor structure
for our dark-matter/quark couplings $c_q$ 
such that
\beq
       c_u = -c_d = c_c = -c_s = c_t = -c_b ~\equiv~  c~.
\label{isospin}
\eeq
We have chosen this flavor structure, which is maximally isospin-violating within each generation,
because it ultimately maximizes the axial-vector decay rate and thereby places the strongest
bounds on our examples.
Second, we shall assume (as a cosmological input) that 
the heavier dark-matter particle $\chi_2$ is metastable and carries 
the vast majority of the dark-matter abundance, \ie,  
$\Omega_2 =\Omega_{\rm CDM}\approx 0.26$ and $\Omega_1 =0$.
As we shall see, this assumption also maximizes the rates for all relevant processes
and thereby places the strongest bounds on our examples.
This assumption also simplifies our analysis somewhat. 
We shall therefore take this to be a conservative ``benchmark'' for our study.
We remark, however, that none of the primary qualitative aspects of our results 
will ultimately depend on this choice, and indeed 
other choices such
as $\Omega_1\approx \Omega_2\approx \Omega_{\rm CDM}/2$
also lead to results which are very 
similar to those we shall obtain here.
Such general scenarios will be explored in Ref.~\cite{toappear}. 

Given these assumptions, our 
examples each have three fundamental parameters:  
the effective coupling $c/\Lambda^2$,
the mass $m_2$ of the heavier dark-matter component, and the dark-sector
mass splitting $\Delta m_{12}\equiv m_2-m_1$.  Our goal is to explore the
resulting $(c/\Lambda^2, \Delta m_{12})$ parameter space for different values of $m_2$.
In this connection, however, 
we remark
that since the operators in Eqs.~(\ref{scalarintpre}) and (\ref{axialintpre})
are non-renormalizable, they can only be interpreted within the context of
an effective field theory whose cutoff scale is parametrically connected to $\Lambda$.
As a result, our use of such operators when calculating phenomenological bounds
already presupposes that the energy scales associated 
with the relevant processes in each case do not exceed $\Lambda$.
Assuming ${\cal O}(1)$ operator coefficients, 
this requires $\Lambda \gsim {\cal O}({\rm GeV})$
for direct-detection bounds;  indeed, as we shall see, 
this value corresponds to nuclear recoil energies $E_R\lsim 100$~keV
in direct-detection experiments. 
Likewise, for indirect-detection bounds, our requirement for $\Lambda$
depends on whether we are dealing with dark-matter annihilation or decay:
for annihilation this requires $\Lambda\gsim {\cal O}(m)$
where $m$ is a typical mass of a dark-sector component, while
for a dark-matter decay of the form $\chi_2\to \chi_1 {\overline{q}} q$ this requires 
$\Lambda \gsim {\cal O}(\Delta m_{12})$.
Finally, for calculating collider-production bounds, the use of such operators
is strictly valid only if $\Lambda \gsim {\cal O}({\rm TeV})$. 
If this last condition is not met, the resulting collider bounds should be viewed only
as heuristic, and one would require a more complete theory (for example, involving potentially light mediators 
connecting the dark and visible sectors) before being able to make more precise statements.

As $\Delta m_{12}\to 0$, our dark-matter production
and direct-detection processes proceed exactly as they would for a single dark-matter
particle $\chi$ of mass $m=m_2$.  Indeed,
in this limit direct detection can only proceed through elastic scattering;
likewise, the interactions in Eq.~(\ref{scalarintpre}) or (\ref{axialintpre})
do not permit dark-matter decay.
Thus, the $\Delta m_{12}\to 0$ limit effectively embodies the physics
of a traditional single-component dark sector with mass $m_2$, 
and in this case our bounds are relatively straightforward:
one simply finds that direct-detection and collider experiments place $m_2$-dependent 
upper limits on the coupling $c/\Lambda^2$. 
For example, for the coupling structure in Eq.~(\ref{isospin}) and
for $m_2= 100$~GeV,
we find from direct-detection 
experiments~\cite{LUX,COUPP4} that
\beq
\begin{cases}
        & {c^{\rm (S)} / \Lambda^2} ~\lsim ~2.8 \times 10^{-10}~{\rm GeV}^{-2}~,\\
        & {c^{\rm (A)} / \Lambda^2} ~\lsim ~1.1 \times 10^{-5}~{\rm GeV}^{-2}~;
 \end{cases}
\label{origlimit}
\eeq
we likewise find from 
collider monojet~\cite{monojetATLAS,monojetCMS}
and mono-$W/Z$~\cite{monoWZATLAS} constraints that 
\beqn
\begin{cases}
         &  {c^{\rm (S,A)} / \Lambda^2} ~\lsim ~1.4 \times 10^{-6}~{\rm GeV}^{-2} ~~ \mbox{(monojet)}~,\\ 
         &  {c^{\rm (S)} / \Lambda^2}   ~\lsim ~5.0 \times 10^{-7}~{\rm GeV}^{-2} ~~ \mbox{(mono-$W/Z$)~,}\\
         & {c^{\rm (A)} / \Lambda^2}   ~\lsim ~3.1 \times 10^{-7}~{\rm GeV}^{-2} ~~ \mbox{(mono-$W/Z$)}~.
\end{cases}
\label{origlimit2}
\eeqn
However, by turning on $\Delta m_{12}$, we can now explore the effects
of non-minimality in the dark sector
and thereby assess the impact of the
new kinematics and new complementarities that arise.

With an eye towards exploring those regions of parameter space 
which are likely to be of 
maximal phenomenological interest
from a complementarity perspective,
we shall restrict our attention to situations with 
$\Delta m_{12}\lsim {\cal O}({\rm MeV})$;  
we stress, however, that there is no fundamental reason
that larger $\Delta m_{12}$ cannot also be considered.
Note that a small mass splitting $\Delta m_{12} \ll m_1, m_2$ can
be realized naturally in models wherein the
generation of $\Delta m_{12}$ is associated with the breaking
of an approximate symmetry (see, \eg, Ref.~\cite{upscattering}).
Examples include models in which a Dirac fermion is split into
a pair of nearly degenerate Majorana states by a small Majorana mass,
 and models in which a complex scalar is split into two real scalars by a small holomorphic mass.

Since we are taking $\Omega_1=0$ for simplicity, 
the relevant dark-matter processes for the operators
in Eqs.~(\ref{scalarintpre}) and (\ref{axialintpre})
are limited to inelastic down-scattering,
asymmetric collider production, 
and dark-matter decay. 
We shall now discuss each of these in turn.

%=========================================================
 {\it  Inelastic down-scattering.}\/--- 
We begin by considering the bounds from direct-detection experiments
through the inelastic down-scattering process $\chi_2 N_i \to \chi_1 N_f$  
where $N_i$ and $N_f$ denote the initial and final states of the
nucleus $N$ in the detector substrate.
In general, the total differential nuclear-recoil rate is given by
\begin{equation}
  \frac{dR}{dE_R} ~=~  \frac{\tilde N \rho_2^{\mathrm{loc}}}{ m_2}
    \int_{v > \vmin^{(21)}} v \mathcal{F}_2(\vec{v})
    \left(\frac{d\sigma_{21}}{dE_R}\right) d^3v~,
\end{equation}
where $\tilde N$ is the number of nuclei per unit detector mass,
where $\rho_2^{\mathrm{loc}}$ is the local energy density
of $\chi_2$, where $\mathcal{F}_2(\vec{v})$ is the distribution of detector-frame
velocities $\vec{v}$ for $\chi_2$ in the local dark-matter halo, where $v=|\vec{v}|$, 
where $\vmin^{(21)}$ is the $E_R$-dependent ``threshold'' velocity (\ie,
the minimum incoming velocity for
which scattering with recoil energy $E_R$ is possible),
and where $d\sigma_{21}/dE_R$ is the differential
cross-section for the process $\chi_2 N_i \rightarrow \chi_1 N_f$.
When evaluating these cross-sections, we
require nuclear form factors;
for this purpose we utilize the software packages associated with Ref.~\cite{Haxton}.
The corresponding limits on our $(c/\Lambda^2,\Delta m_{12})$ parameter
space are then respectively derived using the most recent LUX~\cite{LUX} and COUPP-4~\cite{COUPP4} data 
for the scalar and axial-vector cases, respectively.  
Roughly speaking, this data can be taken as requiring 
$R \lsim  1.81 \times 10^{-4} \, {\rm kg}^{-1} {\rm day}^{-1}$
and $R \lsim  4.97 \times 10^{-2} \, {\rm kg}\/^{-1} {\rm day}^{-1}$
for the recoil-energy windows $3~{\rm keV} \leq  E_R \leq  25~{\rm keV}$
and $7.8~{\rm keV} \leq  E_R \leq  100~{\rm keV}$, respectively.
In this connection, we note that a threshold detector such as COUPP-4
is actually sensitive to scattering events with arbitrarily large recoil
energies.  However, for $E_R\gsim 100$~keV, there are considerable uncertainties associated
with distinguishing these events from background.  To be conservative,
we therefore adopt the above upper limit for the corresponding $E_R$ window.

While much of this analysis is completely standard, the primary new ingredient
is the change in scattering kinematics 
from elastic to inelastic.  
If the scattering had been elastic (\eg, $\chi_j N_i\to \chi_j N_f$),
the possible allowed recoil energies would have been 
given by the standard expression $E_R = E_{jN} (1+ \cos\theta)$ 
where $E_{jN} \equiv \mu_{j N}^2 v_j^2/m_N$ 
and $ 0 \leq \theta \leq \pi$.
Here $v_j$ is the speed of the incoming dark-matter particle
in the detector frame,
$m_N$ the mass of the nucleus,
$\theta$ the scattering angle in the center-of-mass frame,
and $\mu_{\chi N}$ the reduced mass of the $\chi/N$ system.
For inelastic $\chi_j N_i\to \chi_i N_f$ scattering, 
by contrast, the possible allowed recoil energies are 
instead given by $E_R= E_{jN} (1+ r + \sqrt{1+2r} \cos\theta)$
where $r\equiv [\mu_{iN} / ( \beta \mu_{jN})^2 ] \Delta m_{ij}$
with $\beta\equiv v_j/c$ and $\Delta m_{ij}\equiv m_j-m_i$.

%========================= FIGURE BEGINS HERE =====================================
\begin{figure}[b!]
\centerline{
   \epsfxsize 2.2 truein \epsfbox {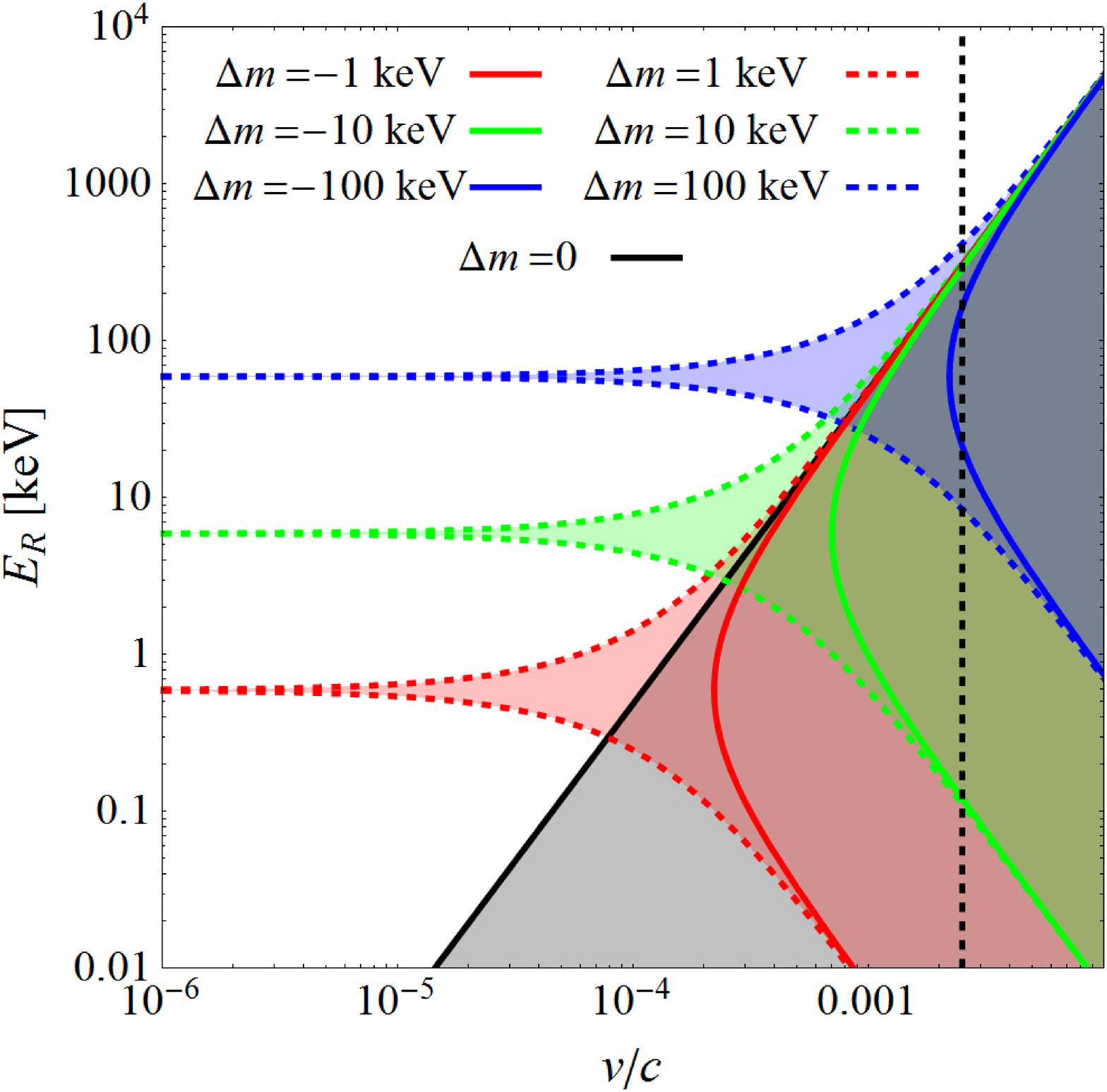}
}
\centerline{
   \epsfxsize 2.2 truein \epsfbox {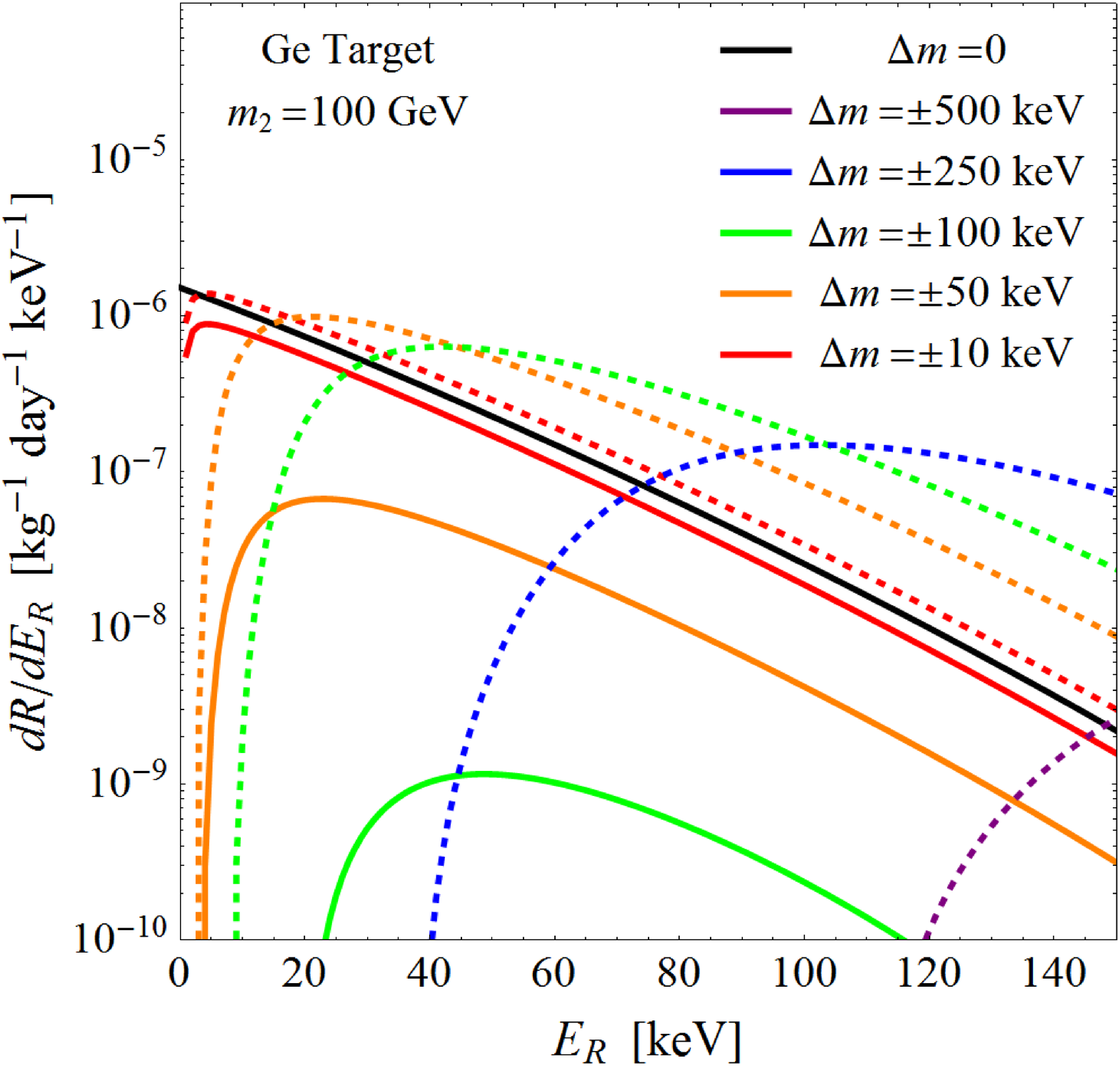} 
      }
\caption{
  Recoil-energy spectra for inelastic scattering $\chi_2 N_i\to \chi_1 N_f$
   off a germanium nucleus, with $m_1\not= m_2=100$~GeV.
 {\it Top panel:}\/  Allowed ranges of recoil energy $E_R$ as a function of   
 incoming dark-matter particle velocity $v$ for different $\Delta m\equiv m_2-m_1$, 
  for both ``down-scattering'' ($\Delta m>0$, dotted lines)
   and ``up-scattering'' ($\Delta m<0$, solid lines).
  The elastic case with $ \Delta m=0$ is also shown (solid black line), as is
   the maximum velocity cutoff
    associated with the galactic escape velocity (dashed black line).
 {\it Bottom panel:}\/  
   Corresponding recoil spectra $dE_R/dR$ for both
        down-scattering (dotted lines)
      and up-scattering (solid lines), for different values of $\Delta m$.
    The solid black curve represents the
  $\Delta m=0$ elastic-scattering case.
   For all spectra shown, the scattering is assumed to be spin-independent, with
  cross-section per nucleon $\sim 10^{-46}\mathrm{~cm}^{-2}$.}
\label{fig:inelastickinematics}
\end{figure}
%=================================================================================

There are two important consequences of this change in kinematics.
First, in the case of down-scattering, we see that 
for $E_R\approx E_R^\ast\equiv [m_i/(m_i+m_N)] \Delta m_{ij} c^2$
the required incoming velocity is essentially zero.
Such threshold-free scattering with non-zero recoil energy 
would not have been possible in the traditional case of elastic
scattering, but the required input energy in the inelastic case
comes directly from species conversion 
within the dark sector (essentially from rest mass liberated within the dark sector)
rather than from incoming dark-sector kinetic energy.  

Second, we also observe that for any incoming dark-matter velocity $v$,
we have not only a finite upper limit on the allowed nuclear recoil energy
$E_R$ but also a non-zero {\it lower}\/ limit.  
This feature holds for both down-scattering and up-scattering,
and in turn implies that 
the corresponding recoil spectrum $dR/dE_R$ is negligible
not only above a maximum value 
of nuclear recoil energy {\it but also below a minimal value.}
Indeed, in the case of down-scattering (\ie, $\Delta m>0$), 
this allowed range of recoil energies is
centered around $E_R^\ast$
and becomes exceedingly narrow as the incoming velocity goes to zero.
This is especially important since
$v\leq v_{\mathrm{esc}} + v_E$, where $v_{\mathrm{esc}} \approx 540$~km/s 
is the galactic escape velocity and $v_E \approx 231$~km/s is the speed of 
the Earth in the galactic frame.
These features are illustrated in Fig.~\ref{fig:inelastickinematics}, and will
play an important role in what follows.

%=====================================================
  {\it Asymmetric collider production.}\/--- 
Multi-component operators such as those in Eqs.~(\ref{scalarintpre}) 
and (\ref{axialintpre}) 
can also be probed through
the collider-production direction.
However, unlike the single-component case, 
the production processes induced by
the operators in Eqs.~(\ref{scalarintpre}) and (\ref{axialintpre}) 
take the form
$qq\to \chi_1\chi_2$.
Thus, we are dealing with {\it asymmetric}\/ collider production of
dark matter.

Despite this new feature,
such dark-matter production processes
can nevertheless be most effectively constrained just
as for single-component theories ---
\ie, through monojet searches at 
ATLAS~\cite{monojetATLAS} and CMS~\cite{monojetCMS} 
and mono-$W/Z$ searches at ATLAS~\cite{monoWZATLAS}.  
Indeed, these limits are directly applicable to the asymmetric production of 
dark-matter particles $\chi_1$ and $\chi_2$ as well as to a pair of identical dark-matter
particles because the values of 
$\Delta m_{12}$ for which inelastic scattering can play a significant role in direct detection 
(and for which $\chi_2$ is stable on cosmologically timescales without exceedingly large $\Lambda$) 
are negligibly small in comparison with the energy scales relevant for collider physics.  
Thus, while the kinematics associated with the asymmetric production of dark matter 
differs from that associated with the more traditional symmetric production,
this difference ultimately does not prove phenomenologically significant within 
the parameter ranges of greatest interest from a complementarity perspective.

%=======================================================
 {\it Dark-matter decay.}\/--- 
We now turn to an analysis of the new (diagonal) complementarity direction which comes into existence 
for non-zero $\Delta m_{ij}$, namely the possibility of decay from heavier to lighter dark-matter
components.
Once again, our starting points are the operators 
in Eqs.~(\ref{scalarintpre}) and (\ref{axialintpre}) which 
describe the microscopic (short-distance) interactions  
between our dark-sector particles $\chi_i$ and Standard-Model quarks.
However, because we are considering dark-sector mass splittings of 
size $\Delta m_{12}\lsim {\cal O}({\rm MeV})$,
our study of the dark-matter decay 
induced by the operators in Eqs.~(\ref{scalarintpre}) and (\ref{axialintpre}) 
must instead be performed
within the framework of a
low-energy {\it macroscopic}\/ effective field theory
in which the physics of 
Eqs.~(\ref{scalarintpre}) and (\ref{axialintpre})
is recast in terms of interactions between
our dark-sector particles and the lightest color-neutral states in the visible sector.

In order to transition between these two descriptions, we can use the formalism
of chiral perturbation theory~\cite{CHPTReviews}
based on the low-energy $SU(2)_L\times SU(2)_R\times U(1)_V$ flavor symmetry group
of the light $(u,d)$ quarks~\cite{GasserLeutwylerO2Loop}.
This technique allows us to systematically generate a complete set
of operators which capture the exact symmetry structure of our underlying microscopic
Lagrangian, up to unknown (but ultimately measurable) coefficients $\tilde \lambda_i\sim {\cal O}(1)$.
The details behind this calculation will be discussed more fully in Ref.~\cite{toappear}.
The upshot, however, is 
that 
the scalar interaction in Eq.~(\ref{scalarintpre}) gives rise to 
an effective Lagrangian of the form~\cite{O6LagNonanomalousFearing}
\beqn
   {\cal L}_{\rm eff}^{\rm (S)}  &=& 
        -{B_0^2 (m_u-m_d) \tilde \lambda_1 c^{\rm (S)} \over 8\pi^2 \Lambda^2 }\, \chibar_2\chi_1\nonumber\\     
      && ~+~ {B_0 \alpha_{\rm EM}\tilde \lambda_2 c^{\rm (S)} \over 4\pi \Lambda_C^2 \Lambda^2}\, 
      (\chibar_2\chi_1) F_{\mu\nu}F^{\mu\nu}~+... ~,~~~~
\label{effscalar}
\eeqn    
where $B_0\equiv m_\pi^2 /(m_u+m_d)$ and where $\Lambda_C$ denotes the confinement scale
$\Lambda_C \equiv 4\pi f_\pi /\sqrt{N_f}$, with $f_\pi \approx 93$~MeV and $N_f=2$.  
Likewise, 
the axial-vector interaction in Eq.~(\ref{axialintpre}) 
gives rise to an 
effective Lagrangian of the form~\cite{O6LagNonanomalousFearing}
\beqn
   {\cal L}_{\rm eff}^{\rm (A)}  &=& 
       - {  c^{\rm (A)} \Lambda_C \over \sqrt{2} \pi \Lambda^2}\left[
      1+ {m_\pi^2 \over 2\Lambda_C^2 } \tilde \lambda_3\right] 
        (\chibar_2 \gamma^\mu \gamma^5 \chi_1)(\partial_\mu \pi^0)\nonumber\\
  && + ~ {\alpha_{\rm EM} \tilde \lambda_4 c^{\rm (A)} \over 4\pi \Lambda_C^2 \Lambda^2}
        (\chibar_2 \gamma^\mu \gamma^5 \chi_1) \partial_\mu (F_{\nu \rho} \tilde F^{\nu\rho}) ~+...~~~~
\label{effaxial}
\eeqn    
We stress, however, that there are numerous subtleties involved in extracting
these terms from the full chiral perturbation theory formalism.
These will be discussed more fully in Ref.~\cite{toappear}.  

Somewhat surprisingly, the first term of Eq.~(\ref{effscalar})
is not an interaction term, but rather an off-diagonal contribution to
the mass matrix of the individual dark-sector components!  
At first glance, this might seem to suggest that the mass eigenstates 
of the dark sector are not given by the individual $\chi_{1,2}$ components 
with which we started, but rather by some new linear combinations of these states.
However, our original supposition has always been that $\chi_{1}$ and $\chi_2$ are 
our physical mass eigenstates with masses $m_1$ and $m_2$ respectively --- 
even in the confined phase of QCD within which we have been operating throughout 
this paper ---  and we know that the mass eigenstates of our theory should not 
change when we reshuffle our strongly interacting degrees of freedom from quarks 
to color-neutral hadrons.  Thus, the contribution from this extra mass term within 
Eq.~(\ref{effscalar}) must ultimately be cancelled by other additional
mass terms (\eg, coming from ultraviolet physics and/or other hadronic effects)
so as to reproduce the mass eigenstates with which we started.  
That said, we might worry whether this cancellation might involve a significant 
degree of fine-tuning.  However, it is easy to verify that this is not the case 
throughout the region of parameter space in which we are primarily interested ---
\ie, that within which $\Delta m_{12} \gsim {\cal O}({\rm keV})$ and
$\Lambda \gsim {\cal O}(10~{\rm GeV})$.  Indeed, within this region of
parameter space, the degree of mixing associated with the mass matrix 
within Eq.~(\ref{effscalar}) is negligible compared to $\Delta m_{12}$.  

We therefore conclude that the mass term appearing within Eq.~(\ref{effscalar}) is
relatively unimportant for the present analysis 
in which we are focusing exclusively on dark-matter phenomenology 
within the confined phase of QCD.  However, it is perhaps nevertheless worth
remarking that terms of this sort can 
potentially play an important role in dark-matter {\it cosmology}\/,
especially if the universe experiences a post-inflationary phase
with a sufficiently high reheating temperature $T>\Lambda_{\rm QCD}$.
At such temperatures, the dark-sector mass matrix could in principle 
differ from its present-day form in a non-trivial manner.  The process 
of confinement itself might then actually {\it induce}\/ a rediagonalization 
of the dark-sector mass eigenstates as part of the QCD phase transition,
precisely through the appearance of terms such as those in Eq.~(\ref{effscalar}).
As a result, the mass eigenstates of the dark sector 
might not be the same before and after the QCD phase transition.  
In general, this is a novel effect which arises only for multi-component 
dark sectors.  However,  this effect can have a wealth 
of important theoretical and phenomenological implications, especially for 
dark-sector components
$\chi_{i}$ and $\chi_j$ whose mass splittings $\Delta m_{ij}$ are extremely small.
These effects will be discussed further in Ref.~\cite{toappear}.

\begin{figure}[t!]
\begin{center}
  \epsfxsize 3.0 truein \epsfbox {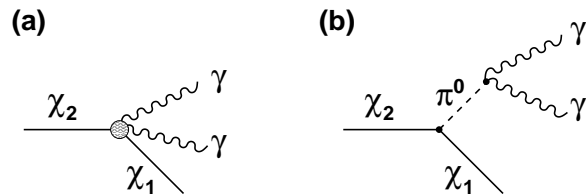}
\end{center}
\caption{Dominant dark-matter decay processes at energies $E \lsim {\cal O}(10^2)~{\rm keV}$.  
(a)  Dark-matter decay produces two photons through an effective contact operator 
induced in the chiral perturbation theory by integrating out heavy hadrons.
This process is the dominant contributor in the case of the microscopic (quark-level) scalar 
interaction in Eq.~(\protect\ref{scalarintpre}).  
(b) Dark-matter decay produces produces two photons via off-shell neutral-pion exchange.
Both this process and the process in (a) are the dominant contributors in the case
of the microscopic (quark-level) axial-vector interaction in Eq.~(\protect\ref{axialintpre}).}
\label{fig:pions}
\end{figure}

The remaining terms in Eqs.~(\ref{effscalar}) and (\ref{effaxial})
are {\it bona fide}\/ interaction terms.  As we see, they come in two 
types:  contact
operators (ultimately obtained from integrating out heavy 
hadronic degrees of freedom) which
directly couple our dark-sector components $\chi_i$ to photons,
and operators which couple our dark-sector components to off-shell pions
(which then subsequently decay to two photons).
These two sets of operators are illustrated in Fig.~\ref{fig:pions}.
Together, however, these operators allow us to calculate the widths
for the decays $\chi_2\to \chi_1 \gamma\gamma$
through either the scalar or the axial-vector interaction.
For $\Delta m_{12} \ll m_{1,2}$, we obtain 
\beqn
   \Gamma_{\gamma\gamma}^{\rm (S)}  &\approx& 
     {B_0^2 \alpha_{\rm EM}^2 \tilde \lambda_2^2 [c^{\rm (S)}]^2 \over
        8(105\pi^5) \Lambda_C^4 \Lambda^4}
         (\Delta m_{12})^7 ~,\nonumber\\
   \Gamma_{\gamma \gamma}^{\rm (A)} &\approx&
       {\alpha_{\rm EM}^2 [c^{\rm (A)}]^2\over 
        8(315 \pi^5) \Lambda_C^4 \Lambda^4}  
             \left[\tilde \lambda_3- \tilde \lambda_4 + {2\Lambda_C^2\over m_\pi^2} \right]^2
        (\Delta m_{12})^9 ~.~~~~~
\label{eq:PartialWidthsToPhotonsApprox}
\eeqn
These decay rates may then be compared against existing bounds
on observed photon fluxes in order to constrain our fundamental parameters
$(c/\Lambda^2, \Delta m_{12})$ for different values of $m_2$.
In particular, assuming an NFW profile for the dark-matter
distribution~\cite{NFW}, we use the PPPC4DMID software package~\cite{Cirelli}
to determine the diffuse galactic and extragalactic contributions to the differential photon flux
arising from dark-matter decay, and require that the predicted photon count
not exceed that measured in any bin at the $2\sigma$ confidence level.
In this context we remark that
decays to final states involving neutrinos are also kinematically allowed. 
However, the contributions to the total widths from such decays are negligible 
compared with the above, and are thus neglected.

%==========================================================================
%==========================================================================

\begin{figure*}[ht!]
\begin{center}
  \epsfxsize 2.0 truein \epsfbox {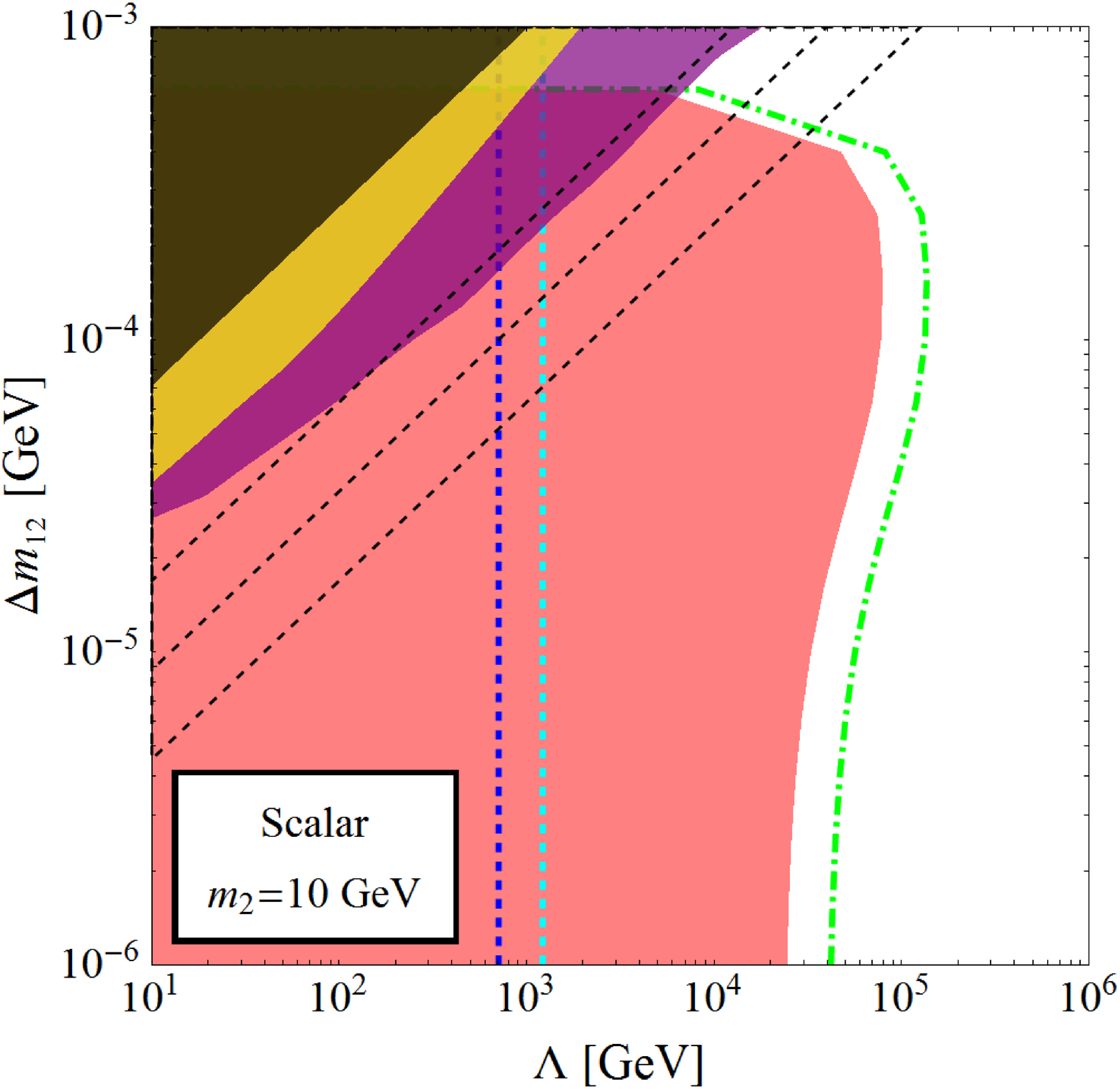}
\hskip 0.2 truein
  \epsfxsize 2.0 truein \epsfbox {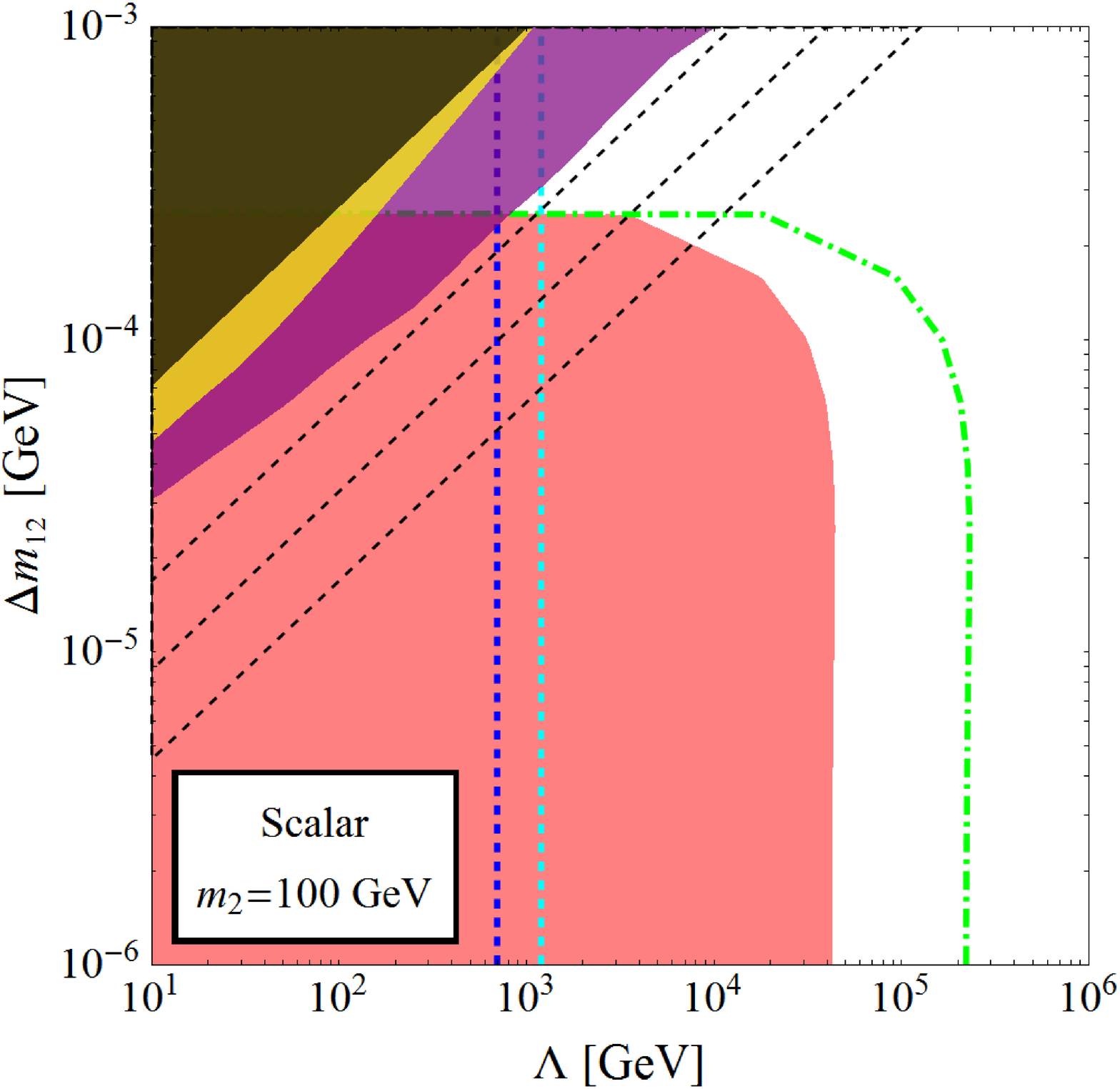}
\hskip 0.2 truein
  \epsfxsize 2.0 truein \epsfbox {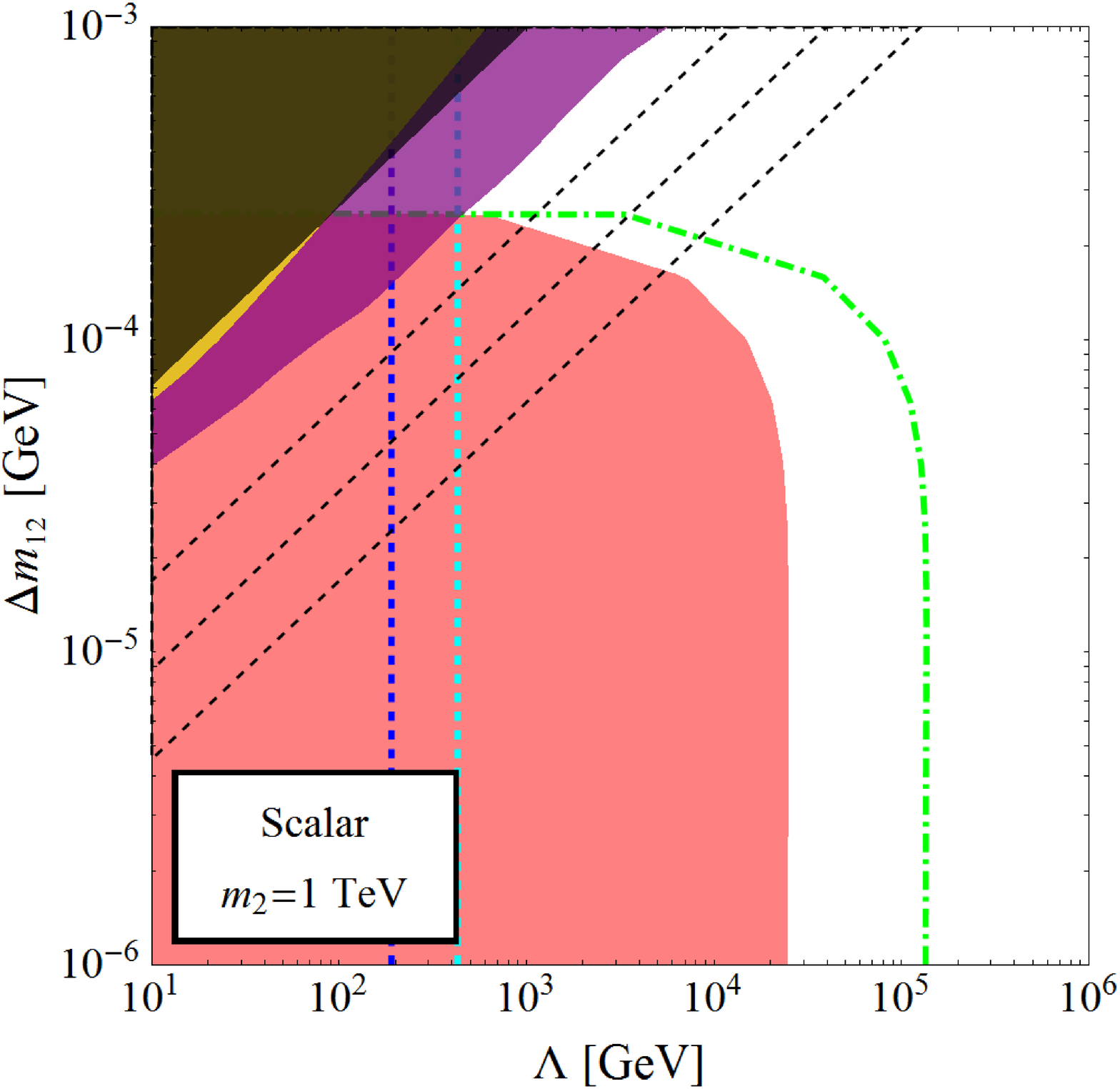}
\end{center}
\begin{center}
  \epsfxsize 2.0 truein \epsfbox {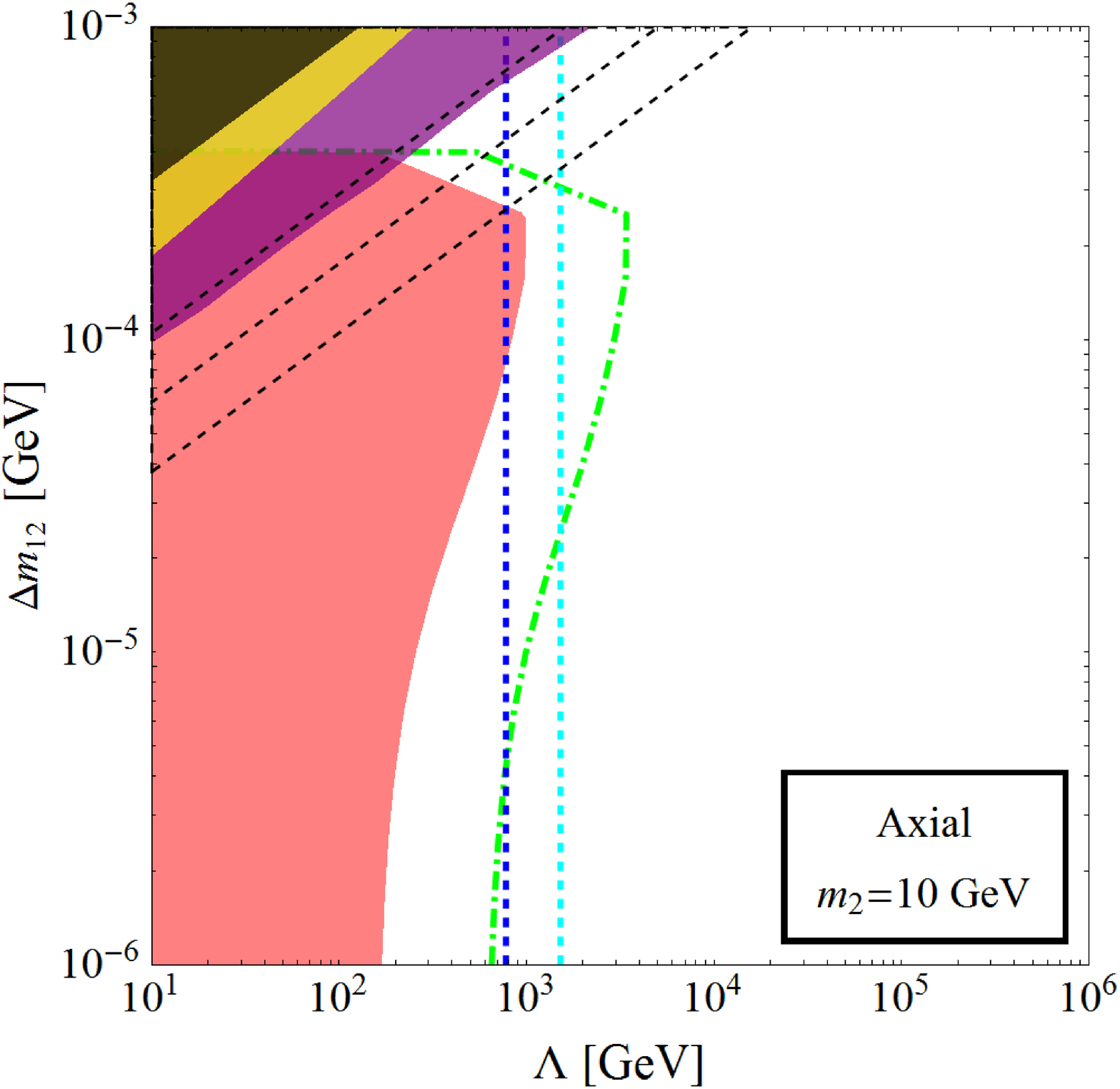}
\hskip 0.2 truein
  \epsfxsize 2.0 truein \epsfbox {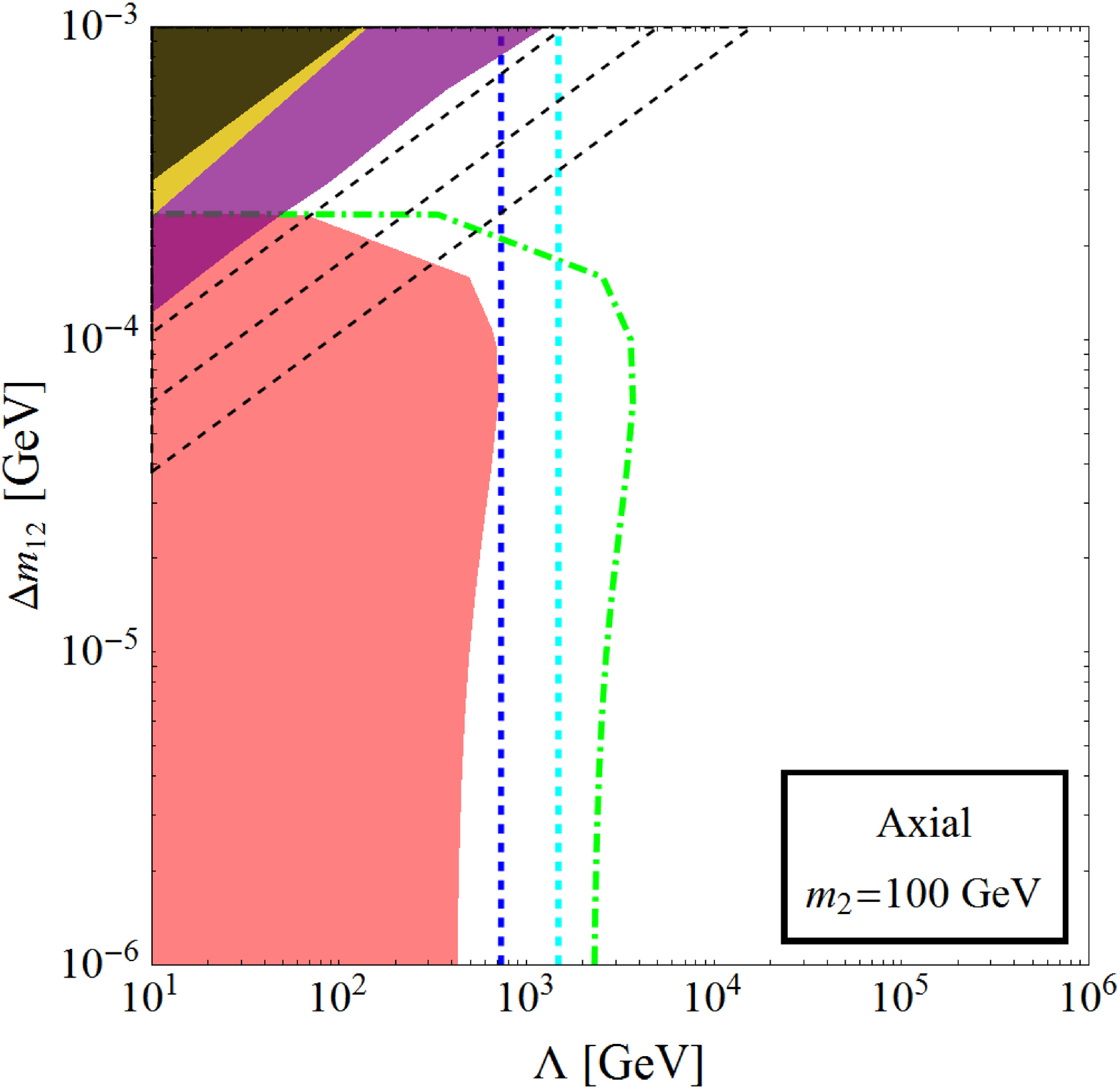}
\hskip 0.2 truein
  \epsfxsize 2.0 truein \epsfbox {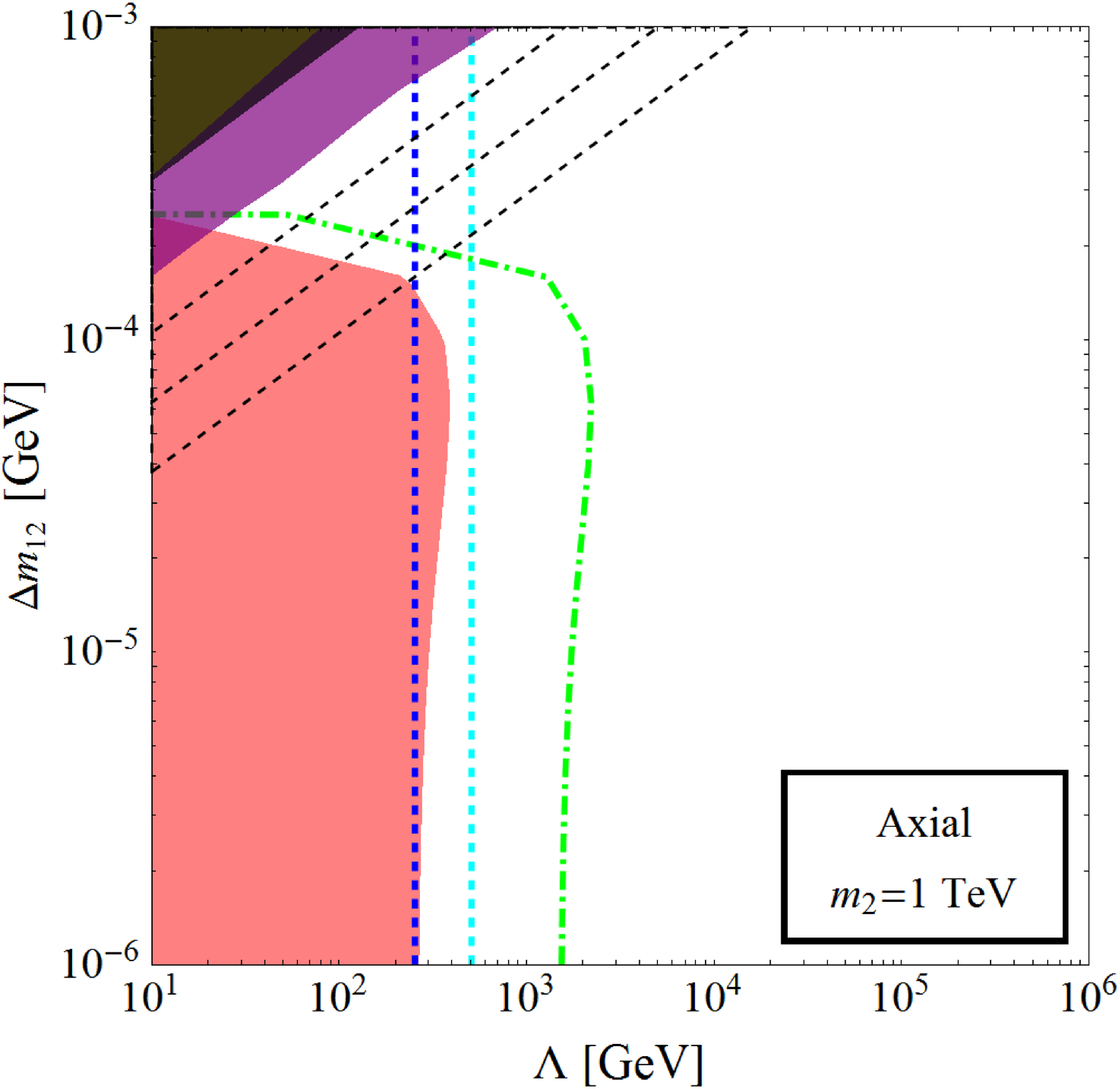}
\end{center}
\caption{
Complementary bounds on the scalar operator in Eq.~(\ref{scalarintpre}) and 
axial-vector operator in Eq.~(\ref{axialintpre}),
plotted within the associated $(\Lambda,\Delta m_{12})$ parameter spaces
for $m_2= \lbrace 10,100,1000\rbrace$~GeV
and $c^{\rm (S)}=c^{\rm (A)}=1/\sqrt{2}$.
Bounds from inelastic-scattering direct-detection experiments (pink exclusion regions), 
asymmetric collider production (blue and cyan vertical lines),
and dark-matter decay constraints (yellow and purple exclusion regions) are shown, as discussed in the text;
the green dashed lines denote the reaches of possible future direct-detection experiments,
while the black dashed lines indicate dark-matter decay lifetime contours, as discussed in the text,
and the solid black triangular regions in each panel are excluded by metastability constraints.
Remarkably, the constraints from dark-matter decay 
dominate in exactly those regions 
with relatively large $\Delta m_{12}$
that lie beyond the reach of current and future direct-detection experiments, 
thereby illustrating the new sorts of complementarities
that are possible for such multi-component dark sectors. } 
\label{results}
\end{figure*}

%==============================================================
{\it Results.}--- 
Combining the constraints from each of the dark-matter directions 
discussed above, we can now 
map out the current bounds on 
the operators in Eqs.~(\ref{scalarintpre}) and (\ref{axialintpre})
in the $(c/\Lambda^2,\Delta m_{12})$ parameter space for
different values of $m_2$.
Our results are shown 
in Fig.~\ref{results} for
$m_2=\lbrace 10,100,1000\rbrace$~GeV, 
with $c^{(\rm S)} = c^{\rm (A)}=1/\sqrt{2}$ and $\tilde \lambda_i=1$ chosen 
as fixed reference values.

%===================== FIGURE STARTS HERE ======================================
%===================== FIGURE ENDS HERE ======================================

In Fig.~\ref{results}, the pink regions are excluded by
bounds on inelastic scattering from direct-detection experiments
(LUX~\cite{LUX} and COUPP-4~\cite{COUPP4} for the scalar and axial-vector
cases, respectively), while the green contour indicates 
the projected future reach of the 
LZ 7.2-ton detector~\cite{LZ} in the scalar case and the PICO-250L experiment~\cite{PICO}
in the axial-vector case.
Likewise, the vertical blue
and cyan contours respectively correspond to LHC constraints on asymmetric collider production  
from monojet~\cite{monojetATLAS,monojetCMS} 
and hadronically-decaying mono-$W/Z$~\cite{monoWZATLAS} searches. 
The collider analysis was performed by generating signal
events using the MadGraph~5~\cite{Alwall:2011uj}, Pythia~6.4~\cite{Sjostrand:2006za}, and Delphes~2.0.5~\cite{Ovyn:2009tx} software packages,
and comparing to the number of background events reported in Refs.~\cite{monojetATLAS,monojetCMS,monoWZATLAS} 
in order to determine the region excluded at $90\%$ confidence level.  Note that there is a 
$\sim 40\%$ systematic uncertainty in the number of
signal events, attributable to uncertainties in 
the correct
treatment of soft QCD and hadronic physics;  this uncertainty
can affect the bounds on $\Lambda$ by $\sim 10\%$. 
In this connection we also remind the reader that these collider-based bounds 
should be interpreted at best only heuristically if the operators in 
Eqs.~(\ref{scalarintpre}) and (\ref{axialintpre})
are generated by integrating out light mediators or by other new
physics which renders $\Lambda \lsim {\cal O}({\rm TeV})$.
The yellow and purple shaded regions are excluded by constraints
on dark-matter decay from the total diffuse X-ray background measurements 
of HEAO-1~\cite{HEAO1} and INTEGRAL~\cite{INTEGRAL}, respectively,
as these are the experiments which probe the particular energy region of the photon
spectrum which is most relevant for the $\Delta m_{12}$ range  
with which we are most concerned.
Finally, the diagonal dashed black lines from left to right 
respectively indicate contours corresponding
to dark-matter lifetimes $\tau_2 = 10^{22}$~s, $10^{24}$~s, and $10^{26}$~s.
By contrast, the solid black triangular regions 
are excluded by metastability constraints which
require that $\tau_2\gsim t_{\rm now}$, where 
$t_{\rm now}\approx 4.35\times 10^{17}$~s 
is the current age of the universe.
Note that in all cases,
these metastability bounds for $\chi_2$ are superseded by
the results from dark-matter decay.
Mechanisms for having such long-lived
dark-matter components can be found, for example, in Ref.~\cite{Fairb}.

There are many important features contained within the plots in Fig.~\ref{results}.
Since $\Delta m_{12}$ effectively 
quantifies departures from the standard single-component story, 
we shall discuss these features ``from bottom up'', 
in order of increasing $\Delta m_{12}$.
   
First, for $\Delta m_{12} \lsim {\cal O}(10~{\rm keV})$,
we see that all of the features within these plots are virtually
insensitive to $\Delta m_{12}$ and effectively
reproduce the physics
of a traditional single-component dark sector with mass $m_2$.
This behavior is certainly expected
in the $\Delta m_{12}\to 0$ limit, and as a check we see that
the middle panels of Fig.~\ref{results}
correctly reproduce the results 
in Eqs.~(\ref{origlimit}) and (\ref{origlimit2}) in this limit. 
However, we now observe that these results 
persist all the way up to 
$\Delta m_{12} \lsim {\cal O}(10~{\rm keV})$, thereby forming
an ``asymptotic'' region in which the physics remains
largely $\Delta m_{12}$-independent.
Thus, for the operators in Eq.~(\ref{scalarintpre})
and (\ref{axialintpre}),
we see that it is only for 
$\Delta m_{12} \gsim {\cal O}(10~{\rm keV})$
that the effects of dark-sector non-minimality become evident.

Second, 
for $\Delta m_{12} \approx {\cal O}(10{-}100~{\rm keV})$,
we see that the bounds from direct-detection experiments actually
strengthen somewhat and begin to extend towards larger values of $\Lambda$.
As we see from Fig.~\ref{results}, this behavior is more pronounced 
for the axial-vector interaction than the scalar interaction
and for smaller values of $m_2$ rather than larger.
This strengthening can more than double the values of $\Lambda$
probed by such experiments,
and is particularly important 
because it has the power to alter
the identity of the specific dark-matter detection method
which provides the leading constraint on the scale $\Lambda$
as a function of $\Delta m_{12}$. 
For example, in the case of the axial-vector
interaction with $m_2=10$~GeV (corresponding to the
lower left panel of Fig.~\ref{results}),
we see that
the monojet and mono-$W/Z$ collider processes provide the strongest constraints
for $\Delta m_{12}\lsim {\cal O}(10~{\rm keV})$, but
that the bounds from the direct-detection processes become increasingly strong
with growing $\Delta m_{12}$, ultimately 
matching and perhaps even superseding the monojet collider bounds
for $\Delta m_{12}\approx {\cal O}(100~{\rm keV})$. 
This strengthening of the direct-detection bounds 
as a function of increasing $\Delta m_{12}$ is a direct
consequence of the inelastic nature of the scattering process involved.

Third, moving towards even larger values $\Delta m_{12}\approx {\cal O}(100-1000~{\rm keV})$,
we see that the plots in Fig.~\ref{results} now reveal a dramatic feature: 
a ``ceiling'' for $\Delta m_{12}$ beyond which
direct-detection experiments cease to provide any bounds at all
and become virtually insensitive to the 
underlying dark-sector physics!
This is also ultimately a consequence of the unique kinematics
associated with inelastic down-scattering.  
As discussed above, 
for down-scattering there is a lower 
limit of nuclear recoil energies 
$E_R^{\rm (min)}$
below which the differential scattering rate $dE_R/dR$ becomes negligible
(see Fig.~\ref{fig:inelastickinematics}).
This lower limit generally increases with increasing $\Delta m_{ij}$
and is not too far below $E_R^\ast$
for the non-relativistic dark-matter velocities concerned.
However, a given dark-matter direct-detection experiment is 
typically designed to probe only a particular window of recoil energies.  
While the precise window of recoil energies depends on the type of experiment and 
the cuts imposed as part of the data analysis, 
this window typically falls within the range $1\mathrm{~keV} \lesssim E_R \lesssim 100\mathrm{~keV}$,
as discussed above.  
Scattering events with recoil energies outside this range do not contribute 
to the measured signal-event rate.
As a result, there exists a critical value of
$\Delta m_{12}$ beyond which the corresponding down-scattering events 
have a minimum recoil
energy $E_R^{\rm (min)}$ exceeding $100$~keV, thereby escaping detection.

Fourth, moving towards even larger values $\Delta m_{12}\lsim {\cal O}({\rm MeV})$, 
we see from Fig.~\ref{results} that the physics is now dominated by the constraints 
from dark-matter decay.
Indeed, these constraints become significantly 
more stringent as $\Delta m_{12}$ increases, 
with the maximum reach $\Lambda_{\rm max}$
scaling approximately as  $(\Delta m_{12})^{7/4}$
and $(\Delta m_{12})^{9/4}$ for the scalar and axial-vector interactions, respectively.
It should also be noted that these decay constraints
even have a subtle dependence on $m_2$.
Indeed, although the decay widths in 
Eq.~(\ref{eq:PartialWidthsToPhotonsApprox})
are independent of $m_2$ for $\Delta m_{12}\ll m_2$,
we see that $m_2$ nevertheless enters the calculation of the total fluxes of the remnants
of such decays through its appearance in the $\chi_2$ number density $n_2\approx 
\Omega_2/m_2$, where $\Omega_2$ is the 
dark-matter abundance which we have assumed fixed at  $\Omega_2 = \Omega_{\rm CDM} \approx 0.26$.
There are, of course, further $m_2$-dependent corrections to both the decay widths and the injection spectra
which emerge once we go beyond the $\Delta m_{12}\ll m_2$ approximation;
these have been included in the plots shown in Fig.~\ref{results}, but otherwise
have a negligible effect on our results.

Finally, although we have restricted our focus in this paper to the region
$\Delta m_{12}\lsim {\cal O}({\rm MeV})$,
it is interesting to contemplate what occurs for even greater $\Delta m_{12}$.
For $\Delta m_{12}\gsim {\cal O}({\rm MeV})$, additional decay channels for $\chi_2$ open up in 
which electron/positron pairs are the end products.
This only increases the decay widths for $\chi_2$, thereby strengthening 
the $\Lambda$-reach of 
the decay-related bounds even further.  
For even greater values of $\Delta m_{12}$,
these decay widths increase still further as additional decay 
channels become kinematically accessible.
Ultimately, however, $\Delta m_{12}$ reaches a point at
which even the metastability
that underlies our assumption of a non-zero present-day abundance for $\chi_2$ is 
threatened.  In this connection, it is important to note that this does not render such
multi-component theories inconsistent;  indeed, for multi-component theories it has
been demonstrated that dark-matter stability is not a fundamental requirement --- all that is
required is a balancing of their individual component lifetimes 
against their cosmological  abundances~\cite{DynamicalDM}.
However, this shift does alter the initial assumptions that enter into the types of
calculations we have performed in this paper.  
It would nevertheless be interesting to extend this type of complementarity analysis 
to the constraints emerging from such a scenario.

The plots in Fig.~\ref{results} thus provide 
dramatic illustration of the new complementarities that emerge 
within the context of non-minimal dark sectors.
Together, the bounds from asymmetric collider-production processes,
inelastic-scattering processes,
and dark-matter decay processes 
not only help to increase the {\it coverage}\/ of
the relevant parameter spaces of these models but 
also provide useful
 {\it correlations}\/ between these processes in those
regions of parameter space in which these constraints overlap.
For example, it is 
somewhat remarkable that the 
constraints from dark-matter decay 
emerge and dominate in exactly those regions 
that lie just beyond the
$\Delta m_{12}$ ``ceiling''
that caps the reach of current and future direct-detection experiments. 
This result is especially gratifying, given that any operators and initial conditions
which give rise to inelastic down-scattering direct-detection signals in
a multi-component context
must also necessarily give rise to dark-matter decays.
Perhaps even more interestingly, we see that in the axial-vector
case there even exists a small window within the ``cross-over'' region 
$\Delta m_{12}\approx 500$~keV (sandwiched between 
the bounds from direct-detection and dark-matter decay processes)
in which it is the mono-$W/Z$ {\it collider}\/ bounds 
which provide the strongest current constraints.
Taken together, this non-trivial structure is testament to 
the richness of the complementarities that emerge when $\Delta m_{12}$ is
lifted beyond the $\Delta m_{12}=0$ axis 
to which the traditional complementarities 
that govern the physics of single-component dark sectors
are restricted.

%======================================================
{\it Conclusions.}\/--- The idea of complementarity has long infused our thinking
about the hunt for dark matter, 
but most work on this subject has focused on the case of
single-component dark sectors.
In this paper, by contrast, we have considered the 
case of a multi-component dark sector,
and demonstrated that there exist entirely new directions for complementarity
which are absent in single-component theories.
In particular, we demonstrated that the important class of interactions
involving two dark components 
and two visible components
can simultaneously contribute to
 {\it inelastic}\/ scattering at direct-detection experiments,
 {\it asymmetric}\/ dark-matter production at colliders,
and indirect-detection signals due to dark-matter {\it decay}\/.
Indeed, the latter phenomenon is completely absent for
such interactions within single-component dark-matter theories, 
and thus represents an entirely
new direction for dark-matter complementarity 
that emerges only within the multi-component context.

We have also demonstrated the power of these complementarity relations
by considering two particular 
examples of such interactions, one based on a scalar (spin-independent) interaction
and the other based on an axial-vector (spin-dependent) interaction.
In some regimes involving large couplings or small cutoff scales $\Lambda$, we found that there is 
significant overlap between
the regions excluded by direct- and indirect-detection limits. 
Taken together, these complementary probes of the dark sector combine to 
provide complete coverage of the relevant parameter space in this regime.
By contrast, in other regimes involving smaller couplings and/or larger cutoff scales $\Lambda$, 
a small slice of parameter space
opens up for which the dark sector escapes detection. 

The existence of such regions of parameter space
provides extra motivation for the development of new 
experimental detection strategies which are specifically 
targetted towards physics in these regions.
For example, it would be interesting to explore
how improvements in, \eg, the angular resolution of
future X-ray telescopes could improve 
the reach of indirect-detection experiments
within the parameter space of non-minimal dark sectors.
Likewise, designing a calibration for incorporating
higher-energy nuclear recoils into threshold-detector analyses 
of the data from direct-detection experiments
also represents a possible future method of ``filling in the gap" 
between the bounds from direct-detection experiments and those from dark-matter decay.
Finally, we note that direct-detection experiments using heavier target nuclei would
in principle be capable of probing regions of parameter space with larger $\Delta m_{12}$.

Needless to say, there are also many future theoretical directions that can be pursued.
One is to consider a wider class of operators beyond those considered here~\cite{toappear}.
The case of pseudoscalar operators, in particular, may be of particular interest
due to the existence of previously unnoticed effects
which are capable of overcoming the velocity suppression 
that would otherwise affect the corresponding direct-detection processes~\cite{pseudoscalar}.

Another possible future direction is to consider 
the physics that might result from different configurations of 
initial abundances $\Omega_1$ and $\Omega_2$.
In this paper we have focused on the case with
$\Omega_1\approx 0$ and $\Omega_2\approx \Omega_{\rm CDM}$, since this configuration
leads to the strongest possible bounds for both direct- and indirect-detection experiments. 
Although this configuration may initially seem somewhat unnatural or fine-tuned,   
one can imagine that it is realized in cosmological scenarios
in which the production of heavier dark-matter states is overwhelmingly 
favored relative to that of lighter dark-matter states, or in which 
the bulk of the dark matter is somehow excited into the higher-mass $\chi_2$ state after production.
It is nevertheless of interest to explore the phenomenology associated with more general configurations,
particularly those such as 
$\Omega_1\approx \Omega_2\approx \Omega_{\rm CDM}/2$ which might be imagined as emerging
from a straightforward thermal-production mechanism.
Obviously, any scenario with non-zero $\Omega_1$ will 
generally involve contributions from processes such as up-scattering (in addition
to down-scattering) and dark-matter co-annihilation (in addition to dark-matter decay).
However, it often turns out that the constraints from both of these processes are subleading within
their respective classes (direct- and indirect-detection signals, respectively);
indeed, co-annihilation 
will not even occur in scenarios (such as those associated with asymmetric dark matter~\cite{asymmDM})
in which the abundance of dark anti-matter does not match that of dark matter
and is effectively zero at present times.
Thus, in such cases, the primary effect of shifting some abundance $\Delta \Omega$ from $\Omega_2$ to $\Omega_1$ 
(thereby resulting in $\Omega_2'\equiv \Omega_2-\Delta \Omega$)
is merely to weaken the constraints from 
down-scattering and dark-matter decay 
by the factor $\Omega_2'/\Omega_2\equiv 1-\Delta \Omega/\Omega_2$. 
On logarithmic plots such as those in Fig.~\ref{results}, such ${\cal O}(1)$ rescaling 
factors are barely noticeable.
Such effects will be discussed further in Ref.~\cite{toappear}.

A third possible future direction is to realize
that even though we have restricted our attention to operators such as those in Eqs.~(\ref{scalarintpre})
and (\ref{axialintpre}) which only couple $\chi_i$ to $\chi_j$ with $i\not=j$, a more general 
theory involving four-fermi operators of this sort is likely to include the ``diagonal'' $i=j$
operators as well.
We did not study such ``diagonal'' operators in this paper because such operators
appear even in single-component theories of dark matter;
they therefore do not represent the new physics we wished to explore.
However, in a general theory, ``diagonal'' and ``non-diagonal'' operators are 
likely to appear together.
In such cases,
as discussed in the Introduction,
both elastic and inelastic scattering events
can simultaneously occur within
a given direct-detection experiment,
while dark-matter production at colliders
can have both symmetric and asymmetric channels
and the cosmic-ray fluxes relevant for
indirect-detection experiments can potentially
include the products of dark-matter self-annihilation
as well as co-annihilation between different dark-matter species
and dark-matter decay.
It will clearly be of interest to study the experimental complementarity bounds that 
emerge when all of these processes are included simultaneously.
In particular, we note that it might even be possible to establish {\it correlations}\/ 
between the signals from dark-matter decay and dark-matter (co-)annihilation in such 
a way as to potentially distinguish these signals from those which
might be produced through other, unrelated astrophysical phenomena such as pulsars.

Finally, a fourth possible direction for future 
study is to recognize that a non-minimal dark sector may have a relatively
large number of individual components which could potentially give rise to collective effects
that transcend the two-component effects studied here. 
A dramatic example of this occurs within the 
so-called ``Dynamical Dark Matter'' (DDM) 
framework~\cite{DynamicalDM};  
this framework 
gives rise to unexpected signatures
not only for collider experiments~\cite{DDMColliders} 
but also for direct-detection experiments~\cite{DDMDirectDet}
and indirect-detection experiments~\cite{DDMIndirect}.
Studying the full 
parameter space of such models,
especially from a complementarity perspective,
should be an interesting exercise~\cite{toappear}.

\begin{acknowledgments}

%%%%%%%%%%%%%%%%%%%%%%%%%%%%%%%%%%%%%%%%%%%%%%%%%%%%%%%%%%%%%%%%%%%%%%%%%%%%%%%%%%%%%%

This work was supported in part by
the U.S.\ Department of Energy under
Grant DE-FG02-13ER-41976 (KRD),
by the U.S.\ National Science Foundation under CAREER Award PHY-1250573 (JK),
by the Natural Sciences and Engineering Research Council of Canada (BT),
and by U.S.\ Department of Energy under
Grant DE-FG02-13ER-42024 (DY).
The opinions and conclusions expressed herein are those of the authors, and
do not represent any funding agency.
We are happy to thank Z.~Chacko and U.~van~Kolck for discussions.

\end{acknowledgments}

%%%%%%%%%%%%%%%%%%%%%%%%%%%%%%%%%%%%%%%%%%%%%%%%%%%%%%%%%%%%%%%%%%%%%%%%%%%%%%%%%%%%%

\end{document}